# MambaUIE: Unraveling the Ocean's Secrets with Only 2.8 GFLOPs

Zhihao Chen[1,[0009-0000-6806-1326]], Yiyuan Ge[1,# [0009-0006-5442-1865]]

[1] Beijing Information Science and Technology University, Beijing, China
2021011561@bistu.edu.cn, geyiyuan@bistu.edu.cn

**Abstract.** Underwater Image Enhancement (UIE) techniques aim to address the problem of underwater image degradation due to light absorption and scattering. In recent years, both Convolution Neural Network (CNN)-based and Transformer-based methods have been widely explored. In addition, combining CNN and Transformer can effectively combine global and local information for enhancement. However, this approach is still affected by the secondary complexity of the Transformer and cannot maximize the performance. Recently, the state-space model (SSM) based architecture Mamba has been proposed, which excels in modeling long distances while maintaining linear complexity. This paper explores the potential of this SSM-based model for UIE from both efficiency and effectiveness perspectives. However, the performance of directly applying Mamba is poor because local fine-grained features, which are crucial for image enhancement, cannot be fully utilized. Specifically, we customize the MambaUIE architecture for efficient UIE. Specifically, we introduce visual state space (VSS) blocks to capture global contextual information at the macro level while mining local information at the micro level. Also, for these two kinds of information, we propose a Dynamic Interaction Block (DIB) and Spatial feed-forward Network (SGFN) for intra-block feature aggregation. MambaUIE is able to efficiently synthesize global and local information and maintains a very small number of parameters with high accuracy. Experiments on UIEB datasets show that our method reduces GFLOPs by 67.4% (2.715G) relative to the SOTA method. To the best of our knowledge, this is the first UIE model constructed based on SSM that breaks the limitation of FLOPs on accuracy in UIE. The official repository of MambaUIE at https://github.com/1024AILab/MambaUIE.

## 1    Introduciton

With the rise of marine ecology research [1-5], the demand for underwater exploration has also risen [6-7]. Images captured by unmanned underwater vehicles (UUVs) are often plagued by severe blurring and color distortion, making it difficult to meet the demands of practical applications. Therefore, researchers have begun to explore how

---
# These authors contributed equally to this work

to apply underwater image enhancement algorithms to platforms such as unmanned underwater vehicles (UUVs). However, due to the limited resources of devices such as unmanned underwater vehicles and underwater cameras, it is difficult for traditional deep-learning models to achieve efficient underwater image enhancement on these platforms [8,9]. Current UIE methods can be mainly classified into physical model-based methods and deep learning-based methods. Physical model-based approaches [10-12] rely on a priori parameters to perform the inverse image degradation process, which needs to be estimated. However, the biggest challenge of this method is the inability to reverse unknown physical processes, which limits its generality and makes it difficult to adapt to dynamic underwater environments.

In recent years, deep learning-based methods have significantly improved the performance of UIE, most of which are based on Convolution Neural Network (CNN) [13-18]. However, CNN-based methods are constrained to local feature extraction and lack the perception of global contextual information. Global information is critical for UIE because objects in real scenes are always intertwined with the background, e.g., overexposure, light scattering, etc., and utilizing only local information limits the performance of UIE methods. To overcome this limitation, [19-21] explored the application of a Vision Transformer (ViT) in UIE to model remote dependencies using ViT, but this approach usually requires significant computational overhead. In addition, there are methods that combine CNN and ViT [6] to achieve better results, but due to the secondary computational complexity of ViT, this limits its upper-performance limit on UIE. Considering that the UIE task is an intensive prediction task, the unmanned underwater vehicle (UUV) needs to perform real-time enhancement of the captured images, which puts stringent requirements on the real-time and deployable line of the UIE method. Therefore, our goal is to explore the limits of efficiency while maintaining the high accuracy of the UIE task.

Recently, state-space modeling (SSM) has gained wide attention. Based on classical SSM research, modern SSM (e.g., Mamba [18]) has been widely applied, such as semantic segmentation [24-27], small target detection [35], and so on. Compared to the evolving ViT architecture, Mamba not only possesses the ability to model long-range dependencies but also exhibits linear complexity in terms of input size. However, when introducing this promising architecture directly into UIE, the model's accuracy cannot match its efficiency. The reason is that UIE places more emphasis on the perception of local fine-grained features than other visual tasks with distinct target features. Unfortunately, Mamba is not good at distinguishing these local fine-grained features.

In this article, we present for the first time an architecture of Mamba dedicated to UIE tasks, called MambaUIE. Compared to the direct adoption of Mamba, our MambaUIE addresses the locality bug and maintains advanced performance while taking full advantage of Mamba's efficiency. Specifically, MambaUIE consists of Dynamic Interaction-Visual State Space Block (DI-VSS) and Spatial Feed-Farword Network (SGFN). Since Visual State Space Block (VSS) focuses on global information modeling, we combine convolution in parallel with VSS to complement Mamba with local features. In order to enhance the integration of the two branches and to integrate global and local information into one feature map, we propose the Dynamic Interaction Block (DIB). It consists of two interaction operations, Spatial Interaction (S-I) and Channel

Interaction (C-I), which exchange information between the two branches. With Spatial Interaction (S-I) and Channel Interaction (C-I), DIB can adaptively re-weight the feature maps of the two branches from either the spatial or channel dimension. Finally, we design an Spatial feed-Farword Network (SGFN) to process the feature maps obtained from DI-VSS and further increase Mamba's local modeling capability. Ultimately, MambaUIE can extract fine-grained visual information in a global view with a very small amount of computation. Compared with other state-of-the-art methods, our proposed method can achieve the unity of efficiency and accuracy. The main contributions of this paper include the following three points:

- To the best of our knowledge, we are the first to successfully apply Mamba to a UIE task. It provides a new benchmark and reference for exploring more efficient UIE in the future.
- This paper presents a novel architecture, MambaUIE, in which we design the Dynamic Interaction-Visual State Space Block to model global dependencies while capturing local fine-grained features.
- The local modeling capability of Mamba is further enhanced by designing Spatial Feed-Farword Network to improve the model's efficiency.

## 2 Related Work

### 2.1 Underwater Image Enhancement Method Based on CNN and Transformer

With the rapid development of convolutional neural networks, more and more researchers are beginning to apply CNNs to underwater data enhancement. Hou et al. [23] proposed an underwater residual convolutional neural network (URCNN) by improving VGG. Wang et al. [24] designed an end-to-end CNN-based network aimed at color correction and deblurring of images. Li et al. [25] created a paired dataset called UIEB and proposed Water-Net as a convolutional neural network of a generalized model. Islam et al. [26] proposed a dataset called UFO-120 and a Deep-SESR model.Deep-SESR is a pure CNN network with a residue-in-residue structure and a multimodal objective function capable of simultaneous image enhancement and super-resolution.

Compared to CNNs, Transformer has shown excellent performance in global modeling. The first Transformer model introduced into the vision domain was ViT [27], and since then, there has been a proliferation of improved methods based on ViT. Peng et al. [21] introduced the ViT module into GAN networks and used it for UIE, where the focus operation is performed channel by channel, directing the network to focus on severely degraded channels. However, convolution in favor of local features is set aside. Ren et al. [6] demonstrated that supplementing local fine-grained features in a ViT block can improve the accuracy of underwater data enhancement by introducing a convolutional module in Swin-ViT. The ViT-based method performs well in global modeling, but the self-awareness mechanism requires secondary computational complexity regarding image size, resulting in a high computational burden.

## 3    Method

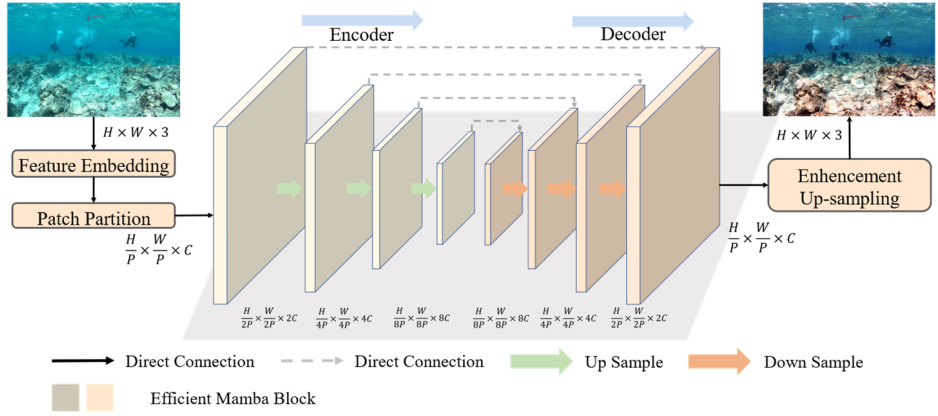

**Fig. 1.** Overall and detailed architecture of the MambaUIE.

### 3.1    Network Architecture

The overall architecture of MambaUIE is shown in Fig. 1. It can be divided into an encoder and a decoder, both of which consist of four stages, each containing an efficient mamba block and upsampling or downsampling. The encoder maps the input to a deeper feature domain while doubling the number of channels and halving the image height and width. The decoder reconstructs the underwater image from the feature domain and increases the image height width while decreasing the number of channels, finally outputting an enhanced underwater image. We remember the original input feature map as $x \in R^{H \times W \times 3}$ and perform feature embedding and patch partition to obtain the feature embedding $F_x \in R^{\frac{H}{P} \times \frac{W}{P} \times C}$ before passing it to the encoder.

**Efficient Mamba Block.** As shown in Fig. 2(1), the core components of an efficient mamba block are a dynamic interaction-visual state space block and an spatial feed-forward network. For the feature embedding $F_x \in R^{\frac{H}{P} \times \frac{W}{P} \times C}$, it is next processed through an encoder consisting of four efficient mamba blocks, where the height and width of the feature map are halved, and the number of channels is doubled after each block. The encoder processes $F_x$ to obtain $F_{en} \in \frac{H}{8P} \times \frac{W}{8P} \times 8C$, which is subsequently passed into the decoder, where the height and width of the feature map are doubled, and the number of channels halved for each efficient mamba block feature map.

*Visual State Space Block:* Fig. 2(3) depicts the visual state space (VSS) block where the input $Map_x$ is divided into two branches after layer normalization. The above process can be represented as:

$$Map_{b1} = Map_{b2} = LN(Map_x) \qquad (7)$$

$LN(\cdot)$ denotes the layer normalization process, and $F_{b1}$ and $F_{b2}$ denote the two branching inputs, respectively. In the first branch, the input is processed through only one linear layer, which can be represented as:

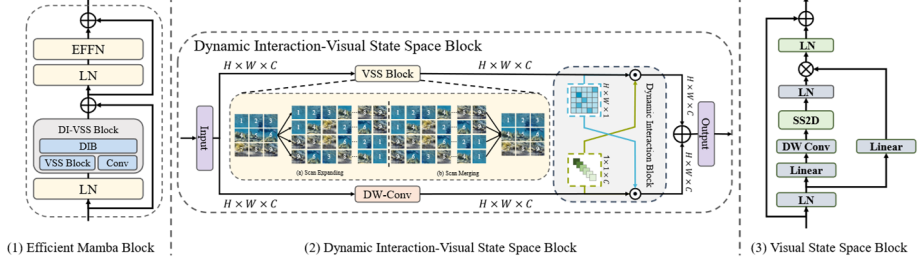

**Fig. 2.** Detailed architecture of the blocks in MambaUIE. (1) shows the core component of MambaUIE. Where (2) shows the detailed structure of Dynamic-Interaction-Visual State Space Block, which consists of the Dynamic Interaction Block, Visual State Space Block (3).

$$Map'_{b1} = Linear(Map_{b1}) \tag{8}$$

Where $Linear(\cdot)$ represents processing using a linear layer. In the second branch, the input is processed with linear layers, depth separable convolution, and activation functions, followed by further feature extraction by applying a 2D selective scanning (SS2D) module. Finally, the features are normalized using layer normalization. The above operations can be represented by the following equation:

$$Map'^{1}_{b2} = DWConv(Linear(Map_{b2}); \theta) \tag{9}$$

$$Map'_{b2} = LN\left(SS2D(Map'^{1}_{b2})\right) \tag{10}$$

$DWConv(\cdot; \theta)$ denotes depth separable convolution and $SS2D(\cdot)$ denotes 2D selective scanning. Then the output of the second branch $Map'_{b2}$ is a matrix dot product with the output of the first branch $Map'^{1}_{b2}$. Finally, it is processed by LayerNorm denoted as $LN(\cdot)$ and residual concatenated with the original input to get the final VSS processed result denoted as $Map_{vss}$:

$$Map_{vss} = Map_x \oplus LN(Map'_{b2} \otimes Map'_{b1}) \tag{11}$$

*Dynamic Interaction Block:* Since the visual state space block focuses on capturing global features, we added a convolution branch parallel to it to introduce local features. However, simply adding a convolutional branch does not effectively combine global and local features. To overcome this problem, we propose the dynamic interaction block (DIB), which acts between the two branches, as shown in Fig. 2(2). The DIB adaptively reweights the features of the two branches in terms of spatial dimensions so that the features of the two branches can be better fused. Based on DIB, we design a new visual state space block called dynamic interaction-visual state space block (DI-VSS).

First, we perform deep convolution (DW-Conv) on the feature embedding $Map_x \in R^{\frac{H}{P} \times \frac{W}{P} \times C}$ to obtain local features of the feature map. We denote the convolution output as $Map_{local}$. Then, we introduce DIB to adaptively adjust the two features, $Map_{local}$ and $Map_{vss}$. Specifically, DIB consists of two interaction operations: global interaction (G-I) and local interaction (L-I). Given two input features $Map_{vss} \in R^{H \times W \times C}$ and $Map_{local} \in R^{H \times W \times C}$, the global interaction computes the global feature map (denoted as $Map_{G\_map}$ with size $R^{H \times W \times 1}$). Local interactions derive the local feature map (denoted as $Map_{L\_map}$) with size $R^{1 \times 1 \times C}$). The operation is shown in Fig. 2(b) and is computed as:

$$Map_{local} = DWConv(Map_x; \theta) \qquad (12)$$

$$Map_{G\_map} = f(\varphi PWConv(Map_{vss}; \theta)) \qquad (13)$$

$$Map_{L\_map} = f(\varphi PWConv(GP(Map_{local}); \theta)) \qquad (14)$$

Where $DWConv(\cdot; \theta)$ denotes deep convolution, $PWConv(\cdot; \theta)$ denotes pointwise convolution, $\theta$ is the process parameter; $\varphi$ is the GELU activation function, $f$ is the sigmoid function, and $GP(\cdot)$ denotes global mean pooling. Subsequently, the global and local feature maps are applied to the corresponding branches, thus enabling interaction to obtain $Map_G$ and $Map_L$ and feature fusion. This process can be formulated as:

$$Map_G = Map_{G\_map} \odot Map_{local} \qquad (15)$$

$$Map_L = Map_{L\_map} \odot Map_{vss} \qquad (16)$$

$$Map_{fusion} = Map_G \oplus Map_L \qquad (17)$$

Finally, using DIB, we designed a Dynamic Interaction-Visual State Space Block based on Visual State Space Block.

## 4 Experiments

### 4.1 Datasets and Evaluation Metrics

We evaluated the model on the UIEB dataset[8], which includes real-world underwater images grouped into two subsets: 890 pairs of raw underwater images with the corresponding high-quality reference images and 60 challenging images without reference. Then, 800 pairs of original images and clear images were extracted from UIEB to train the model. The remaining 90 images in UIEB named T90 were used to test the effect of our method on degraded images. In order to acquire quantitative measurements, we use Peak Signal-to-Noise Ratio (PSNR) [44], Structural Similarity Index (SSIM) [45], as performance metrics for image quality. PSNR is a full-reference image quality evaluation metric based on errors between corresponding pixels. The higher the PSNR score, the better the image quality. SSIM measures the visual quality of three features of an image: brightness, contrast, and structure. A higher SSIM value indicates a higher similarity between the enhanced and reference images.

### 4.2 Implement Details

The proposed MambaUIE is implemented by Pytorch 2.0.0 with an NVIDIA L20 GPU without pretrained networks. The Adam optimizer with the initial learning rate 5e−4 and $\beta \in (0.9, 0.999)$ is utilized for the training, processing 800 epochs with the batch size 8. Besides, the cosine annealing learning rate decay strategy [46] is synergized for training with warm-up epoch 3.

### 4.3 Comparison with the state-of-the-art methods

We compared MambaUIE with several state-of-the-art methods, including traditional methods and deep learning methods. Traditional methods included UDCP [8], IBLA [37], SMBL [41], and MLLE [56], and deep learning methods included UWCNN [28], Water-Net [27], PRWNet [16], Shallow-uwnet [34], Ucolor [29], UIEC^2-Net [47], UHD-SFNet [49], PUIE-Net [10] and the latest NU2Net [11] for underwater image enhancement. We present the comparison of the objective metrics with previous SOTA methods in Table 1. From that, we can observe that our method achieves the best results on PSNR and SSIM, proving that the proposed architecture has good results with detailed textures, restoring promising contrast and color of images. Compared with the last method, NU2Net on T90, we exceed 3.001dB and 0.031 on PSNR and SSIM, respectively.

Table 1: Experimental results on T90 datasets. ↑ represents the higher is the better as well as ↓ represents the lower is the better. The efficiency evaluation uses 720P images as input on L20 GPU.

| Methods | T90 | | |
|---|---|---|---|
| | PSNR↑ | SSIM↑ | GFLOPs |
| UDCP(ICCVW'13)[8] | 13.415 | 0.749 | - |
| IBLA(TIP'17) [37] | 18.054 | 0.808 | - |
| WaterNet(TIP'19) [27] | 16.305 | 0.797 | 193.7 |
| SMBL(TB'20)[41] | 16.681 | 0.801 | - |
| UWCNN(PR *20)[28] | 17.949 | 0.847 | - |
| PRW-Net(ICCVW'21)[16] | 20.787 | 0.823 | 223.4 |
| Shallow-uwnet(AAAI'21)[34] | 18.278 | 0.855 | 304.75 |
| Ucolor(TIP'21) [29] | 21.093 | 0.872 | 443.85 |
| UIEC^2-Net(SPIC*21)[47] | 22.958 | 0.907 | 367.53 |
| MLLE(TIP'22) [56] | 19.561 | 0.845 | - |
| UHD-SFNet(ACCV'22)[49] | 18.877 | 0.810 | 15.24 |
| PUIE-Net(ECCV'22)[10] | 21.382 | 0.882 | 423.05 |
| NU2Net(AAAI'23,Oral) [11] | 22.419 | 0.923 | 146.64 |
| Ours | **25.42** | **0.954** | **2.715** |

## 5      Conclusion

In this paper, we introduce an SSM-based UIE model, MambaUIE. Specifically, we design efficient mamba blocks to introduce visual state space (VSS) blocks at the macro level to capture global contextual information while mining local information at the micro level. Also, for these two kinds of information, we propose a Dynamic Interaction Block (DIB) and Spatial Feed-Forward Network (SGFN) to achieve intra-block feature aggregation. Finally, we experimentally verify the potential of the proposed method in UIE tasks, and MambaTSR achieves SOTA performance with only 2.8 GFLOPs in UIEB datasets, breaking the limitation of FLOPs on accuracy in UIE.

Contribution Title (shortened if too long) 9